\documentclass[12pt]{article}
\pdfoutput=1
\usepackage{amsmath,amsthm,amssymb}
\usepackage{wasysym}
\usepackage{graphicx}
\usepackage[doublespacing]{setspace}
\usepackage{lipsum}
\usepackage[lmargin=2.5cm,rmargin=2.5cm,tmargin=3cm,bmargin=3cm]{geometry}
\usepackage{multirow}
\usepackage{hyperref}
\usepackage{cite}
\usepackage{cleveref}
\usepackage[numbers,sort&compress]{natbib}
\begin{document}
\begin{center}
{\rm \bf \Large{An asperity-based statistical model for the adhesive friction of elastic nominally flat rough contact interfaces}}
\end{center}

\begin{center}
{\bf Yang Xu$^{a}$\footnote{Corresponding author, Yang.Xu@hfut.edu.cn}, Julien Scheibert$^b$, Nikolaj Gadegaard$^c$, Daniel M. Mulvihill$^{d}$\footnote{Corresponding author, Daniel.Mulvihill@glasgow.ac.uk}}\\
{
$^a$School of Mechanical Engineering, Hefei University of Technology, Hefei, 230009, China\\
$^b$Univ Lyon, Ecole Centrale de Lyon, ENISE, ENTPE, CNRS, Laboratoire de Tribologie et Dynamique des Syst\`{e}mes LTDS, UMR 5513, F-69134, Ecully, France\\
$^c$Division of Biomedical Engineering, James Watt School of Engineering, University of Glasgow, Glasgow, G12 8LT, UK\\
$^d$Materials and Manufacturing Research Group, James Watt School of Engineering, University of Glasgow, Glasgow, G12 8QQ, UK
}
\end{center}

\begin{center}
{\bf Abstract}
\end{center}
Contact mechanics-based models for the friction of nominally flat rough surfaces have not been able to adequately capture certain key experimentally observed phenomenona, such as the transition from a static friction peak to a lower level of sliding friction and the shear-induced contact area reduction that has been observed in the pre-sliding regime especially for soft materials. Here, we propose a statistical model based on physically-rooted contact mechanics laws describing the micromechanics of individual junctions. The model considers the quasi-static tangential loading, up to full sliding, of the contact between a smooth rigid flat surface and a nominally flat linear elastic rough surface comprising random independent spherical asperities, and accounts for the coupling between adhesion and friction at the micro-junction level. The model qualitatively reproduces both the macroscopic shear-induced contact area reduction and, remarkably, the static friction peak without the need to explicitly introduce two different friction levels. It also demonstrates how the static friction peak and contact area evolution depend on the normal load and certain key microscale interface properties such as surface energy, mode mixity and frictional shear strength. “Tougher” interfaces (i.e. with larger surface energy and smaller mode mixity parameter) are shown to result in a larger real contact area and a more pronounced static friction peak. Overall, this work provides important insights about how key microscale properties operating at the asperity level can combine with the surface statistics to reproduce important macroscopic responses observed in rough frictional soft contact experiments.

\vspace{5pt}


\section{Introduction}
The tribological properties (including contact, friction and adhesion) of rough contact interfaces between elastic solids have been widely investigated in the last decades (see e.g. \cite{Vakis18} for a recent review). In particular, they are of interest to a wide range of applications involving soft materials such as rubber (including tires \cite{Tolpekina19} and seals \cite{Vladescu19}) or biological materials (including human skin \cite{Sahli18, Huloux21} and animal adhesive pads \cite{Kamperman10, Bullock08}). The mechanical response of rough elastic contact interfaces under pure normal load, in particular the amount of real contact area of the many micro-junctions formed between the highest antagonist asperities, is now rather well understood (see e.g. \cite{Muser17} for a detailed comparison between the many modeling methods available in the literature). In contrast, much less is known about the behaviour of those interfaces under shear. Recent measurements have shown that, well before the onset of macroscopic sliding, significant modifications to the morphology of the contact occurs, including both a reduction in real contact area \cite{Sahli18, Weber19} and a growth in anisotropy \cite{Sahli19}. These shear-induced changes potentially affect all of the macroscopic properties of the interface (including friction, stiffnesses and conductivities) compared to what they are under pure normal loading. It is thus paramount to develop contact and friction models that incorporate shear-induced contact morphology changes. Building and investigating the behavior of one such improved model is the main objective of the present work.

To the authors knowledge, there has been only one attempt in this direction in the literature: that of Scheibert \emph{et al}. \cite{Scheibert20}. They proposed a dynamic independent-asperity model aimed at reproducing quantitatively the experimental results obtained in \cite{Sahli18}. The model is initiated with a list of micro-junctions, the number and individual areas of which are extracted from the experiments, when no shear is applied yet. The shear-induced anisotropic area reduction of each micro-junction is then computed based on empirical laws inspired by the experiments. The slipping threshold at the micro-junction level is assumed to be proportional to the micro-junction area. Overall, most model parameters are directly taken from the experiments. As a result, while the macroscopic tangential force evolution (including the initial shear stiffness and the final stick-slip regime) was well reproduced, the area evolution was not. Specifically, the model predicted a linear area decay as a function of tangential load, instead of the quadratic-like decay observed experimentally. The reasons invoked for such discrepancy were (i) that the empirical microscale behavior laws may be too simplistic to accurately represent the experimental evolution and (ii) a possible role of the elastic interactions between micro-junctions that were not accounted for in their model. As we will see, our model will enable testing of the plausibility of the first of the two above-mentioned reasons, by using one of the most recent adhesive contact models as a microscale law.

To model the onset of sliding of soft rough contact interfaces, our strategy is the following. We will use a statistical model based on independent elastic micro-junctions, as classically done in normal contact \cite{Greenwood66, Fuller75, Greenwood17} or friction models \cite{Braun08, Thogersen14}. At the individual micro-junction scale, we then need two behavior laws. First, a contact mechanics model describing the effects on the micro-junction area of both the normal indentation and the shear displacement. Second, a friction law describing the tangential force sustained by a micro-junction when it is in a slipping state. For the latter law, many models use Amonton’s law of friction with a friction force proportional to the local normal load, from the classical Cattaneo-Mindlin model \cite{Popov17, Prevost13} to dynamic multi-asperity contact models \cite{Troborg14, Tromborg15}, through to models based on the phenomenological rate- and state-dependent friction law \cite{Li20} . Here, we will use the classical law of a friction force proportional to the contact area with the proportionality being the frictional shear strength, a law found generally valid for soft rough contacts \cite{Wu10a, Deg12, Yashima15, Sahli18}. Concerning the contact mechanics model, we will classically assume that the summits of the asperities of the rough surfaces are paraboloids, and thus use the most relevant contact models from the literature.
In this respect, the fact that soft parabolic contacts shrink under shear has been repeatedly observed \cite{Sahli19, Petitet08, Waters10, Mergel19, Lengiewicz20} since the pioneering work of Savkoor and Briggs \cite{Savkoor77}. A recent model \cite{Lengiewicz20} showed that, at least for millimeter-sized contacts, shear-induced area reduction is a non-linear elastic effect due to large deformations in soft materials induced by their low elastic modulus, when involved in an interface with a high frictional shear strength. For smaller contacts, and \emph{a fortiori} for micro-junctions, the prevailing view is that adhesion plays an important role. Thus, in the present work, we will neglect the potential effects of finite strains on the contact and friction response of rough contacts, and concentrate on the role of adhesion. Most of the modeling literature on shear-induced contact area reduction concerns adhesive parabolic contacts, treated either analytically (see e.g. \cite{Savkoor77, Waters10, Papangelo19a, Papangelo19b, Das20, Peng21} or using finite elements (see e.g. \cite{Mergel19, Salehani19, Mergel21}). For our model, it is more convenient and computationally inexpensive to resort to an analytical model. Among those available, we decided to use the one developed by Papangelo and Ciavarella \cite{Papangelo19a}, because they have identified the model ingredients suitable for a quantitative agreement with the experimental results \cite{Mergel19}, which have been obtained on the same materials and tribometer as those used here \cite{Sahli18, Sahli19}.

The manuscript is organized as follows. In Section 2, we describe the adhesive friction model used at the micro-junction level, and present its main features. In Section 3, we present our statistical multi-asperity contact model and demonstrate its general qualitative behavior in Section 4. Although the model parameter values used in the illustrations of section 4 are inspired by actual experimental observations, only in section 5 do we address the challenging question of how to extract the relevant topographical and material parameters enabling a quantitative comparison with experiments. Indeed, in section 5, we apply our model to the conditions of the experiments of Sahli \emph{et al}. \cite{Sahli18}, quantitatively compare the predictions with the observations, and discuss the outcome with respect to that of the previous modeling attempt of \cite{Scheibert20}.

\section{Single asperity adhesive friction model}
Consider a contact pair between an elastic axisymmetric parabolic asperity of radius $R$ and a rigid flat, subjected to a combined action of normal load $P$ and tangential load $T$. The Young's modulus and Poisson's ratio of the asperity are $E$ and $\nu$, respectively. The reduced elastic modulus is $E^* = E/(1 - \nu^2)$. In the rest of this article, Tabor's parameter, $\mu = \displaystyle \left( \frac{R w^2}{E^{*2} \epsilon^3} \right)^{1/3}$, with $w$ being the work of adhesion and $\epsilon \approx 0.2$ nm as the equilibrium distance between atoms, is assumed to be larger than $5$ so that the adhesive contact is within the Johnson, Kendall and Roberts (JKR) limit \cite{Johnson97}.

\subsection{Purely normal adhesive contact: JKR theory}
Because $\mu > 5$, the single spherical adhesive contact under purely normal load (when $T = 0$) is governed by the JKR theory \cite{Johnson71}. Following either the thermodynamic approach \cite{Johnson71} or the fracture mechanics approach \cite{Greenwood81, Maugis92}, the final forms of the JKR theory may be written as follows:
\begin{align}
\delta &= \frac{a^2}{R} - \sqrt{\frac{2 \pi w a}{E^*}}, \label{E:JKR_indent} \\
     P &= \frac{4 E^* a^3}{3 R} - \sqrt{8 \pi E^* a^3 w}, \label{E:JKR_load}
\end{align}
where $a$ and $\delta$ are the radius of the circular contact area and the indentation depth, respectively. Note that only the stable branch is given above. As the work of adhesion vanishes ($w \to 0$), the JKR theory reduces to the Hertzian theory \cite{Johnson87}.

In asperity-based rough surface adhesive contact models (e.g., the Fuller-Tabor model \cite{Fuller75} and the Greenwood model \cite{Greenwood17}), single asperity contact is in fixed grip condition (i.e., indentation is prescribed). Thus, the unstable points of the JKR theory in fixed grip condition at loading and unloading stages are discussed below.

{\bf Loading stage} Due to the intermolecular attractions (e.g., van der Waals forces) between two mating surfaces, two approaching asperities can immediately jump onto contact with an initial gap, which is known as the jump-to-contact distance, that can be represented using a negative indentation: $\delta_{\text{JKR}}^{\text{loading}} < 0$. Because the original JKR theory does not account for the adhesion outside the contact area, it is impossible to derive the jump-to-contact distance using the JKR theory. Thus, a zero jump-to-contact distance is commonly assumed when using the JKR theory \cite{Greenwood17}. Using a numerical modeling approach, an empirical solution of $\delta_{\text{JKR}}^{\text{loading}}$ was fitted by Wu \cite{Wu10}:
\begin{equation}\label{E:JKR_delta_min_loading}
  \delta_{\text{JKR}}^{\text{loading}} = -2.641 \mu^{3/7} \exp\left( -\frac{1}{2 \sqrt{\mu}}\right) \epsilon.
\end{equation}

{\bf Unloading stage} The instability (snap-out) also occurs when the contacting asperities are unloaded. The jump-off-contact distance is represented by a negative indentation depth: $\delta_{\text{JKR}}^{\text{unloading}} < 0$. This distance can be directly obtained from the JKR theory \cite{Maugis92}:
\begin{equation}\label{E:JKR_delta_min_unloading}
  \delta_{\text{JKR}}^{\text{unloading}} = -\frac{3}{4} \pi^{2/3} \mu \epsilon.
\end{equation}
The jump-off-contact and jump-to-contact distances ($\delta_{\text{JKR}}^{\text{unloading}}$ and $\delta_{\text{JKR}}^{\text{loading}}$) of single asperity contact are different, which is one of the key factors contributing to the adhesive hysteresis (energy dissipation) of the adhesive rough surface contact in loading-unloading cycles \cite{Greenwood17, Popov21}. Because $\mu > 5$, we can confirm numerically that $\delta_{\text{JKR}}^{\text{unloading}} < \delta_{\text{JKR}}^{\text{loading}} < 0$.

\subsection{Papangelo and Ciavarella (PC) adhesive friction model}
After the indentation has reached the target value during the purely normal loading stage, the contacting asperities are then subjected to a quasi-statically increasing tangential displacement $\delta_T$. Before the contact enters the sliding stage, its periphery can be modeled as a bi-material interfacial crack in a mixed mode loading. The past experimental studies of interfacial cracks \cite{Cao89, Reeder90, Benzeggagh96} showed an increase of the interfacial toughness with phase angle of loading attributed to the energy dissipation through viscoelastic deformation and interfacial friction, etc. Some phenomenological relations have been proposed \cite{Hutchinson90, Hutchinson92} to approximate the measured failure locus and one of them was used by Papangelo and Ciavarella \cite{Papangelo19a} in their single asperity adhesive friction model. In order to maintain a stick stage in PC's model, the following fracture criterion must be satisfied at the contact edge (crack tip)\cite{Papangelo19a}:
\begin{equation}\label{E:failure_locus}
\frac{1}{2 E^*} \left( K_{\text{I}}^2 + K_{\text{II}}^2 \right)= w f(\psi),
\end{equation}
where $K_{\text{I}}$ and $K_{\text{II}}$ are the Mode I (opening) and Mode II (shearing) stress intensity factors (SIFs), respectively; $\psi = \tan^{-1}(K_{\text{II}}/K_{\text{I}})$ represents the phase angle of the loading; $f(\psi) = \left[ 1 + (\lambda - 1) \sin^2(\psi)\right]^{-1}$; $\lambda \in [0, 1]$ is the mode mixity parameter. The left-hand side of Eq. \eqref{E:failure_locus} is the total strain energy release rate at the  contact edge, $G$, which is commonly referred to as the driving ``force" of smearing the adhesive bond. It is clear from the increasing nature of $f(\psi)$ in Eq. \eqref{E:failure_locus} that an interface under mixed mode requires a larger driving force to propagate the interface crack, and that this driving force increases with tangential loading (i.e., with the phase angle $\psi$).

When $\lambda = 1$, the fracture criterion in Eq. \eqref{E:failure_locus} becomes the Griffith criterion ($G = w$), and PC's model reduces to the seminal Savkoor and Briggs adhesive friction model \cite{Savkoor77} where strengthening of the bonding toughness under mixed mode loading is ignored. When $\lambda = 0$, Eq. \eqref{E:failure_locus} becomes $\displaystyle{\frac{1}{2 E^*} K_{\text{I}}^2 = w}$ which indicates that the failure of adhesive bonding is only related to the normal load: the external work due to the tangential loading is completely dissipated along the interface. Therefore, we can conclude that, the smaller the mode mixity parameter $\lambda$, the higher the resistance of the contact to tangential loading.

In Ref. \cite{Papangelo19a}, the PC model is formulated in terms of normal and tangential loads ($P, T)$. For a given pair of normal and tangential displacements $(\delta, \delta_T)$, an alternative form of PC's model may be given as follows:
\begin{align}
\delta &= \frac{a^2}{R} - \sqrt{ \frac{2 \pi a w}{E^*} - \frac{4}{9} \lambda \delta_T^2 }, \label{E:PC_indent} \\
P &= \frac{4 E^* a^3}{3 R} - \sqrt{8 \pi E^* a^3 w -  \frac{16}{9} \lambda E^{*2} a^2 \delta_T^2}, \label{E:PC_P} \\
T &= \frac{4}{3} E^* a \delta_T. \label{E:PC_T}
\end{align}
Note that $\nu = 0.5$ and only the stable branch of the solution is given above. As $\lambda \to 0$, Eqs. \eqref{E:PC_indent} and \eqref{E:PC_P} reduce to the JKR theory. The extra terms in Eqs. \eqref{E:PC_indent} and \eqref{E:PC_P} introduce a coupling between tangential and normal loadings, i.e., the tangential displacement $\delta_T$ influences the normal load $P$ for a prescribed indentation $\delta$. The extent of the coupling is governed by the mode mixity parameter $\lambda$.

\begin{figure}[h!]
  \centering
  \includegraphics[width=\linewidth]{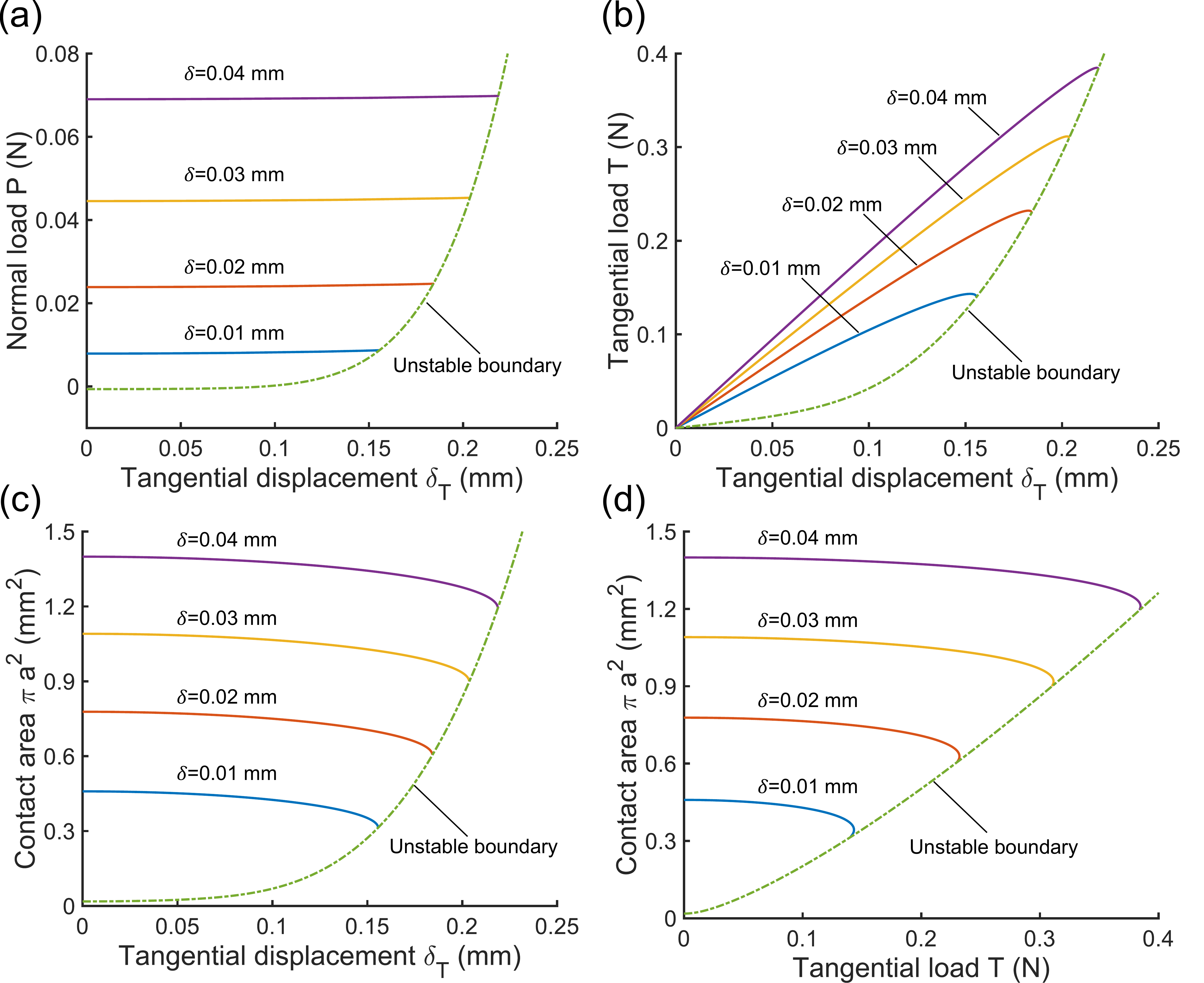}
  \caption{Evolutions of (a) normal load, $P(\delta_T)$, (b) tangential load, $T(\delta_T)$, (c) contact area, $A(\delta_T) = \pi a^2$, and (d) $A(T)$ predicted by PC's model under fixed grip condition for various $\delta$. The model behaviour is illustrated for the contact parameters of \cite{Mergel19}, as extracted by Papangelo and Ciavarella \cite{Papangelo19a}: $w = 27$ mJ/m$^2$, $R = 9.42$ mm, $E^* = 2.133$ MPa, and $\lambda = 0.0023$. Tabor's parameter $\mu \approx 5736$.\label{fig:Fig_1}}
\end{figure}

Using as an example the parameters of the smooth, milimeter-scaled ball-on-flat adhesive frictional test done by Mergel \emph{et al.} \cite{Mergel19}, $P(\delta_T)$, $T(\delta_T)$ and $A(\delta_T) = \pi a^2$ predicted by PC's model are given in Fig. \ref{fig:Fig_1}(a-c) assuming the ball-on-flat contact is under fixed grip condition. After the ball is pressed normally with a constant indentation $\delta$, a monotonically increasing tangential displacement $\delta_T$ is applied. The normal load $P$ slightly increases with respect to the tangential displacement, see Fig. \ref{fig:Fig_1}(a). This minor change might be due to the loose coupling between normal and tangential loading ($\lambda = 0.0023$). The tangential load, $T$, first increases then decreases (this trend is observable for the smallest indentation value, e.g. for the curve associated with $\delta = 0.01$ mm in Fig. \ref{fig:Fig_1}(b). The contact area in Fig. \ref{fig:Fig_1}(c) shows a monotonic decreasing trend which is consistent with the junction shrinkage phenomenon observed in many friction tests \cite{Savkoor77, Waters10, Sahli18, Mergel19, Das20, Lengiewicz20}. The curves $P(\delta_T)$, $T(\delta_T)$ and $A(\delta_T)$ stop at the unstable boundary (for fixed grips condition), beyond which no real root can be found from Eqs. (\ref{E:PC_indent}--\ref{E:PC_T}). From a physical point of view, this is equivalent to saying that the contacting asperities subjected to any tangential displacement $\delta_T > \delta_T^{\text{max}}$ can no longer maintain the stick condition. The sliding stage is assumed to occur beyond this unstable point, even though there is a lack of empirical evidence.

For a given indentation value $\delta$, the unstable points on curves $A(\delta_T)$ are associated with infinite tangent slopes or $\displaystyle \frac{d \delta_T}{d a} \biggl |_{a_\text{min}} = 0$, where $\delta$ is a constant and the corresponding radius has the minimum value $a_\text{min}$. Thus, a cubic equation of $a_\text{min}$ is derived:
\begin{equation}\label{E:PC_a_min_1}
a_{\min}^3 - \delta R a_{\min} - \frac{\pi w R^2}{2 E^*}= 0.
\end{equation}
Equation \eqref{E:PC_a_min_1} results in a unique real positive root as long as the JKR model with the same $(\delta, R, E^*, w)$ has a stable solution. Substituting $a_\text{min}$ into Eq. \eqref{E:PC_indent}, we can get the maximum tangential displacement allowed in the stick stage:
\begin{equation}\label{E:max_disp_PC}
  \delta_T^{max} = \sqrt{\frac{9}{4 \lambda} \left[ \frac{2 \pi w a_\text{min}}{E^*} - \left( \frac{a_\text{min}^2}{R} - \delta \right)^2 \right]}.
\end{equation}
The unstable boundary predicted by Eq. \eqref{E:max_disp_PC} is plotted as a dash-dotted line in Fig. \ref{fig:Fig_1}(a-d).

Similarly, for a given $\delta_T$, we can find the minimum indentation, $\delta_{\text{PC}}^{\text{min}}$, below which the sliding stage starts immediately. Letting $\displaystyle \frac{d \delta}{d a}\biggl |_{a'_\text{min}} = 0$, we can get another cubic equation:
\begin{equation}\label{E:Instable_boundary}
\frac{2 \pi w}{E^*} \left( a'_\text{min} \right)^3 - \frac{4}{9} \lambda \delta_T^2 \left( a'_{\text{min}} \right)^2 - \left( \frac{\pi w R}{2 E^*} \right)^2 = 0.
\end{equation}
The minimum indentation $\delta_{\text{PC}}^{\text{min}}$ can be obtained by substituting the unique real root $a'_{\text{min}}$ of Eq. \eqref{E:Instable_boundary} and $\delta_T$ into Eq. \eqref{E:PC_indent}. It is easy to show that $\delta_{\text{PC}}^{\text{min}} = \delta_{\text{JKR}}^{\text{unloading}}$ when $\delta_T = 0$, and Fig. \ref{fig:Fig_2}(a) indicates that  $\delta_{\text{PC}}^{\text{min}} > \delta_{\text{JKR}}^{\text{unloading}}$ where $\delta_T > 0$. This minimum indentation is important in the modeling of rough surface adhesive friction contact.

\subsection{Transitions between stick, sliding and out-of-contact states}
In this section, a status map is generated to determine the status of an asperity adhesive contact subjected to a certain combination of normal and tangential displacements $(\delta, \delta_T)$. To make the status map more general, the dimensionless normal displacement $\tilde{\delta}$, tangential displacement $\tilde{\delta}_T$, and contact radius $\tilde{a}$ suggested by Papangelo and Ciavarella \cite{Papangelo19a} are slightly modified as follows:
\begin{equation*}
  \xi = \left( \frac{E^* R}{w} \right)^{1/3}, ~\tilde{\delta} = \frac{\xi^2}{R} \delta,  ~\tilde{\delta}_T = \frac{2 \xi^2}{3 R} \delta_T, ~\tilde{a} = \frac{\xi}{R} a.
\end{equation*}
The unstable boundary, $\tilde{\delta}=\tilde{\delta}_{\min}^{\text{PC}}(\tilde{\delta}_T)$, for the single asperity adhesive contact has the following dimensionless form:
\begin{equation}
\tilde{\delta}_{\min}^{\text{PC}} = \tilde{a}_{\min}^2 - \sqrt{2 \pi \tilde{a}_{\min} - \lambda \tilde{\delta}_T^2},
\end{equation}
where $\tilde{a}_{\min}$ is the unique positive root of the dimensionless form of Eq. \eqref{E:Instable_boundary}:
\begin{equation}
\tilde{a}_{\min}^3 - \frac{\lambda}{2 \pi} \tilde{\delta}_T^2 \tilde{a}_{\min}^2 - \frac{\pi}{8} = 0.
\end{equation}
\begin{figure}[h!]
  \centering
  \includegraphics[width=\textwidth]{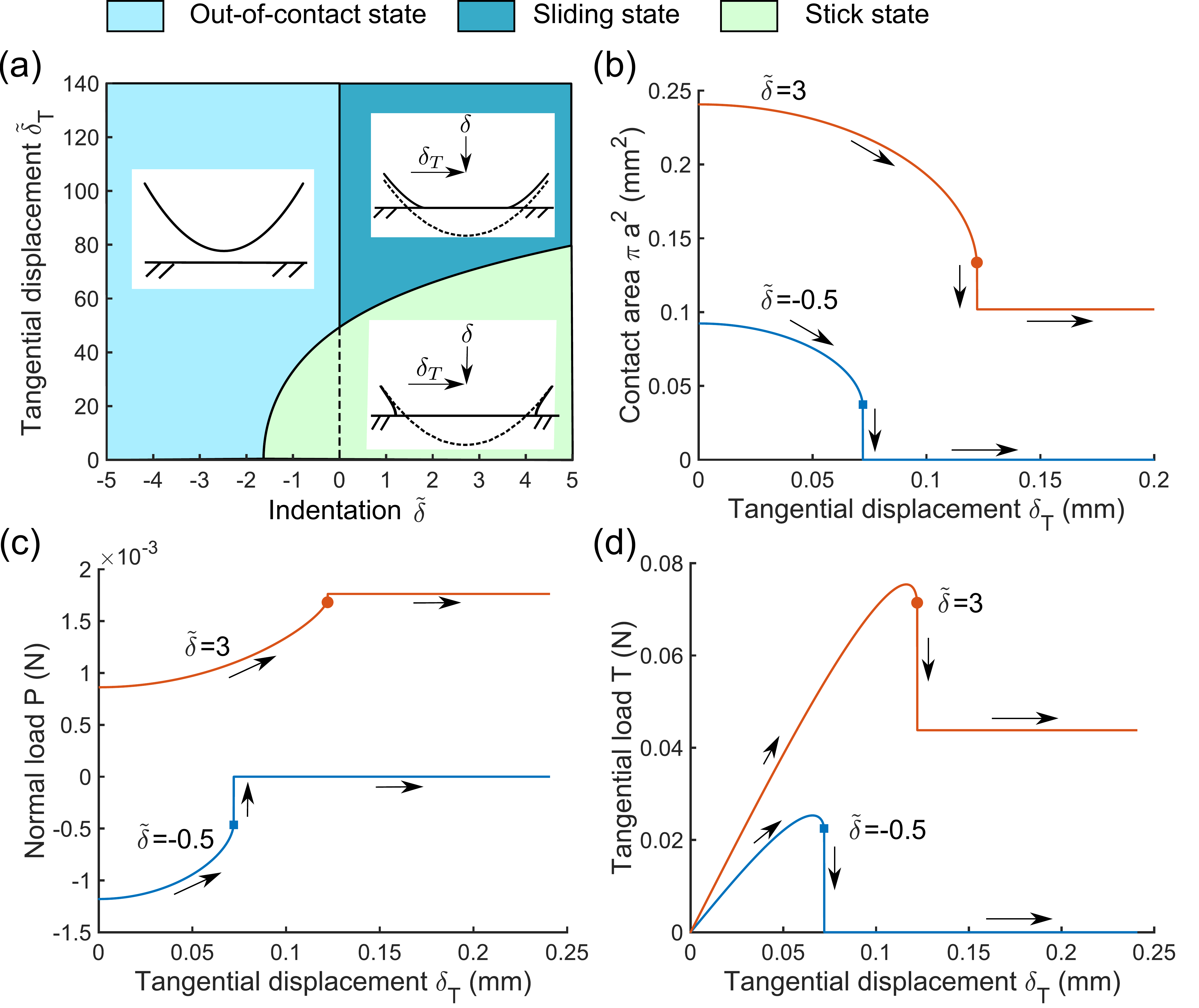}
  \caption{(a) Adhesive friction status map of a single asperity contact subjected to both normal and tangential displacements $(\tilde{\delta}, \tilde{\delta}_T)$. (b, c, d) Evolutions of the contact area $\pi a^2$, normal load $P$, and tangential load $T$, respectively, as functions of the tangential displacement, $\delta_T$, at fixed dimensionless indentations of $\tilde{\delta} = -0.5$ (blue) or $3$ (red). The model behaviour is illustrated for the contact parameters of \cite{Mergel19}, as extracted by Papangelo and Ciavarella \cite{Papangelo19a}: $w = 27$ mJ/m$^2$, $R = 9.42$ mm, $E^* = 2.133$ MPa, $\lambda = 0.0023$, and $\tau_0 = 0.43$ MPa. Tabor's parameter $\mu \approx 5736$.\label{fig:Fig_2}}
\end{figure}

The boundaries, $\tilde{\delta} = \tilde{\delta}_{\min}^{\text{PC}}(\tilde{\delta}_T)$ and $\tilde{\delta} = 0$, split the entire $(\tilde{\delta}, \tilde{\delta}_T)$ domain into three different states, namely, out-of-contact, sliding and stick states, see Fig. \ref{fig:Fig_2}(a). For a constant indentation $\tilde{\delta} < \tilde{\delta}_{\text{JKR}}^{\text{unloading}} = \displaystyle{-\frac{3}{4} \pi^{2/3}}$, which is the jump-off-contact distance in the JKR theory, the asperities are in the out-of-contact state regardless of any value of tangential displacement. If a constant indentation $\tilde{\delta}$ is maintained within $[\displaystyle{-\frac{3}{4} \pi^{2/3}}, 0]$, then the asperity contact will transit from the stick state to the out-of-contact state when $\tilde{\delta}_T$ reaches $\tilde{\delta}_T^{max}$. This phenomenon was confirmed experimentally by Waters and Guduru \cite{Waters10} and Mergel \emph{et al.}, \cite{Mergel19}. The corresponding evolutions $A(\delta_T)$, $P(\delta_T)$, and $T(\delta_T)$ associated with $\tilde{\delta} = -0.5$ are plotted as blue curves in Fig. \ref{fig:Fig_2}(b-d). If the indentation $\delta$ is held at a positive level, asperities will transit from the stick state to the sliding state at the unstable boundary $\tilde{\delta} = \tilde{\delta}_{\min}^{\text{PC}}(\tilde{\delta}_T)$. The sliding state is governed by the Hertzian contact where the shear stress is assumed to be uniform along the contact area with a fixed magnitude of $\tau_0$. The contact area, normal load and tangential load are determined only by the normal loading:
\begin{equation}\label{E:Hertz_theory}
a = \sqrt{R \delta}, ~~~ P = \frac{4}{3} E^* \sqrt{R} \delta^{3/2}, ~~~ T = \pi a^2 \tau_0.
\end{equation}
Note that the transition from stick to slip at the single asperity level is associated with a drop of tangential load, from a peak value accounted for by PC’s model, down to a slipping plateau at the value given in Eq.(14). The evolutions $A(\delta_T)$, $P(\delta_T)$, and $T(\delta_T)$ associated with $\tilde{\delta} = 3$ are plotted as red curves in Fig. \ref{fig:Fig_2}(b-d). The establishment of the status map in Fig. \ref{fig:Fig_2}(a) is helpful for the development of the rough surface friction model in the next section.

\section{Rough surface contact: a statistical model}\label{sec:model}
\begin{figure}[h!]
  \centering
  \includegraphics[width=\textwidth]{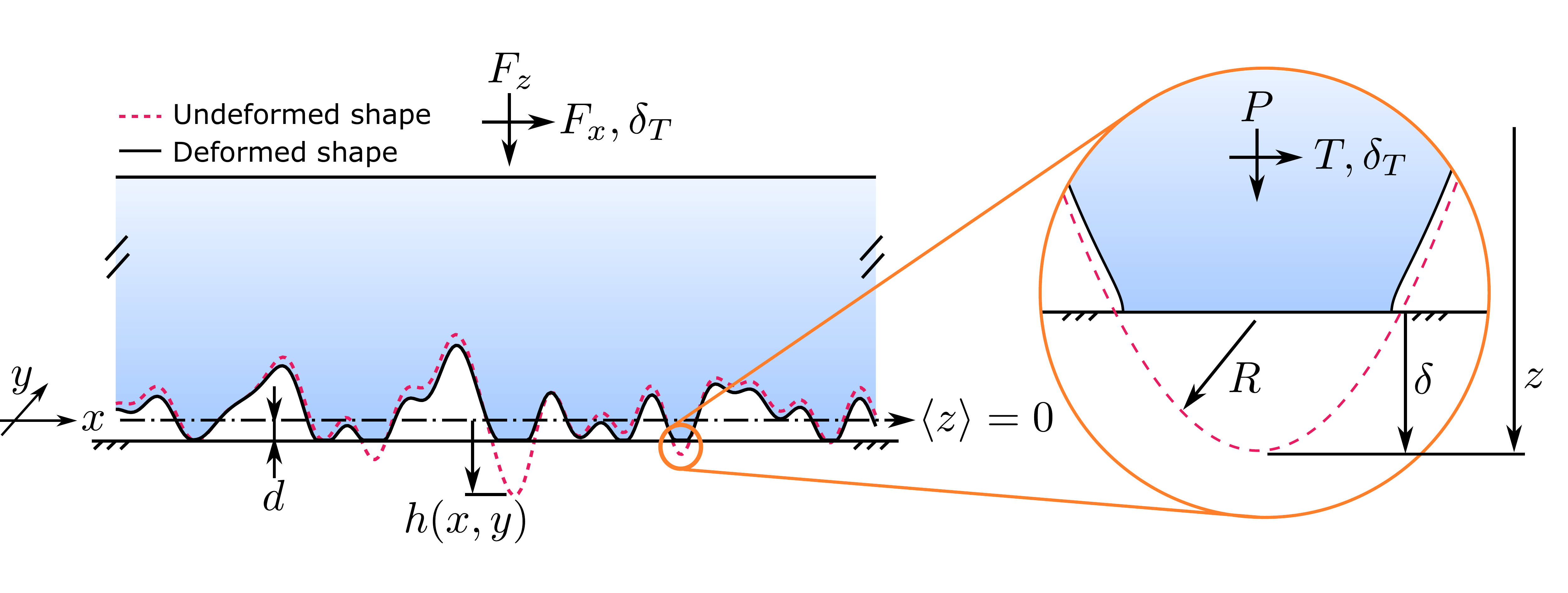}
  \caption{Schematic of the contact interface, loading configuration and associated parameters considered in the statistical adhesive friction model, at both the macroscopic rough surface and individual asperity levels.\label{fig:Fig_3}}
\end{figure}

Consider the specific case where one linear elastic $(E, \nu)$ nominally flat rough surface is in normal contact with a smooth rigid flat over a nominal contact area of $A_n$, see Fig. \ref{fig:Fig_3}. The rough surface topography is denoted by $h(x,y)$. The two surfaces are initially compressed under a purely normal load $F_z$ acting remotely. Then, a tangential displacement $\delta_T$ is applied to the contact pair in a quasi-static manner.

In this section, a statistical model of independent, adhesive spherical asperities is developed. This new model can simulate the adhesive friction, $F_x$, and real contact area, $A_r$, under the combined action of the normal load $F_z$ and of the tangential displacement $\delta_T$. The introduction of this statistical model is divided into two parts: namely, the pre-loading stage and the tangential loading stage.

\subsection{Pre-loading stage}\label{subsec:pre-loading stage}
During the pre-loading stage, the normal load, $F_z$, is monotonically increased from zero and held at a constant value of $F_z^\text{max}$, with the tangential displacement $\delta_T$ kept at zero. A statistical model recently developed by Greenwood \cite{Greenwood17}, based the classic Fuller-Tabor (FT) model \cite{Fuller75} is used. In the FT model, the summits of all asperities have a common radius of curvature, $R$, and the summit height, $z$, follows the Gaussian distribution, i.e.,  $\displaystyle \phi(z) = \frac{1}{\sqrt{2 \pi \sigma_s^2}} \exp \left( -\frac{z^2}{2 \sigma_s^2}\right)$ where $\sigma_s = \sqrt{\langle z^2 \rangle}$ is the standard deviation of asperity summits heights (the mean summit height $\langle z \rangle = 0$). For a given surface separation, $d$, between the rigid flat and the mean summit height (see Fig. \ref{fig:Fig_3}), we can identify all possible contacting asperities based on the value of indentation $\delta = z - d$. Neglecting the coalescence and interaction of neighboring asperities, the normal load, $F_z(d)$, and the real contact area, $A_r(d)$, can be given in the integral forms \cite{Fuller75, Greenwood17}:
\begin{align}
F_z (d) &= \eta A_n \int_{\delta_{\text{JKR}}^{\text{loading}}}^{\infty} P_{\text{JKR}}(\delta) \phi(d + \delta) d \delta, \label{E:Fy_Greenwood} \\
A_r (d) &= \eta A_n \int_{\delta_{\text{JKR}}^{\text{loading}}}^{\infty} \pi a_{\text{JKR}}^2(\delta) \phi(d + \delta) d \delta, \label{E:Ar_Greenwood}
\end{align}
where $\eta$ is the asperity density over the entire nominal contact area $A_n$. The single asperity contact model is governed by the JKR theory. Therefore, for a given $\delta$, the contact radius $a_{\text{JKR}}$ and normal load $P_{\text{JKR}}$ at the asperity level can be solved from Eq. \eqref{E:JKR_indent} and \eqref{E:JKR_load}, respectively. Greenwood \cite{Greenwood17} improved the original Fuller-Tabor model in the loading stage by introducing the jump-to-contact distance $\delta_{\text{JKR}}^{\text{loading}}$, see Eq. \eqref{E:JKR_delta_min_loading}. In the original Fuller-Tabor model, $\delta_{\text{JKR}}^{\text{loading}} = 0$.

As the normal load reaches its maximum, $F_z \to F_z^{\max}$, the corresponding surface separation monotonically reduces to its minimum value $d \to d_{\min}$. Note that if a surface is described by its full topograph $h(x, y)$, one needs to evaluate the roughness-related inputs of the statistical model (i.e., $\eta$, $\sigma_s$ and $R$). One possible method to achieve such an evaluation is described and briefly discussed in Appendix A.

\subsection{Tangential loading stage}
Once entering the tangential loading stage, each contacting asperity initially in the stick state is subjected to normal and tangential displacements $(\delta, \delta_T)$, simultaneously. Based on the status map in Fig. \ref{fig:Fig_3}(a), contacting asperities may transit to an out-of-contact or sliding state, depending on the values of $(\tilde{\delta}, \tilde{\delta}_T)$. Because of the coupling introduced by the mixed mode parameter $\lambda$, the normal load acting on each contacting asperity increases with the tangential displacement, see Fig. \ref{fig:Fig_2}(c) for an example. In order to maintain the macroscopic load equilibrium of rough surface contact in the normal direction, the corresponding surface separation must increase from $d_{\min}$ as tangential displacement proceeds. As a matter of fact, during shear, all contacting asperities tend to unload along the normal to the contact.

Inspired by the loading-unloading formulation in Greenwood's model \cite{Greenwood17}, the corresponding normal load, tangential load and real contact area may be formulated in the following integral forms:
\begin{align}
  F_z(d, \delta_T) = &\eta A_n \int_{\delta_2}^{\infty} P_{\text{PC}}(\delta, \delta_T) \phi(d + \delta) d\delta ~ + \eta A_n \int_{\delta_1}^{\delta_2} P_{\text{Hertz}}(\delta) \phi(d + \delta) d \delta, \label{E:NewFormulation_Fy} \\
  F_x(d, \delta_T) = &\eta A_n \int_{\delta_2}^{\infty} T_{\text{PC}}(\delta, \delta_T) \phi(d + \delta) d\delta ~ + \eta A_n \int_{\delta_1}^{\delta_2} T_{\text{Hertz}}(\delta) \phi(d + \delta) d \delta, \label{E:NewFormulation_Fx} \\
  A_r(d, \delta_T) = &\eta A_n \int_{\delta_2}^{\infty} \pi a_{\text{PC}}^2(\delta, \delta_T) \phi(d + \delta) d\delta ~ + \eta A_n \int_{\delta_1}^{\delta_2} \pi a_{\text{Hertz}}^2(\delta) \phi(d + \delta) d \delta, \label{E:NewFormulation_Ar}
\end{align}
where the two limits $\delta_1$ and $\delta_2$ have the following composite forms:
\begin{align}
\delta_2 &= \text{max}(d_{\text{min}} - d + \delta_{\text{JKR}}^{\text{loading}}, \delta_{\text{PC}}^{\text{min}}), \label{E:delta_2} \\
\delta_1 &= \text{min}(0, \delta_2) \label{E:delta_1}.
\end{align}
The first integrals on the right-hand sides of Eqs. (\ref{E:NewFormulation_Fy}-\ref{E:NewFormulation_Ar}) are associated with all asperities in the stick state where the normal load $P_{\text{PC}}$, tangential load $T_{\text{PC}}$, and contact radius $a_{\text{PC}}$ at the single asperity level are governed by the PC model, see Eqs. (\ref{E:PC_indent}-\ref{E:PC_T}). The second integrals are associated with the asperities in the sliding state, where $P_{\text{Hertz}}$, $T_{\text{Hertz}}$, and $a_{\text{Hertz}}$ are governed by the Hertzian theory, see Eq. \eqref{E:Hertz_theory}.

The composite form of $\delta_2$ is built upon a similar form proposed by Greenwood \cite{Greenwood17}. The lower limit $\delta_2$ is not simply $\delta_{\text{PC}}^{\min}$ \cite{Greenwood17}, because at the end of the loading stage with $\delta_T = 0$, the minimum indentation is $\min(\delta) = \delta_{\text{JKR}}^{\text{loading}} > \delta_{\text{JKR}}^{\text{unloading}} = \delta_{\text{PC}}^{\min}$. Because of this composite form, the lower limit $\delta_2$ will gradually converge to $\delta_{\text{PC}}^{\min}$ as the separation distance $d$ increases. The composite form of $\delta_1$ is determined based on the left boundary of the Hertzian contact zone in the status map of Fig. \ref{fig:Fig_2}(a).

To prove the continuity of $F_z$, $F_x$ and $A_r$ at the transition point between the pre-loading and tangential loading stage, let $\delta_T = 0$ and $d = d_{\min}$, then $\delta_{\text{PC}}^{\text{min}} \equiv \delta_{\text{JKR}}^{\text{unloading}}$, $\delta_2 = \max(\delta_{\text{JKR}}^{\text{loading}}, \delta_{\text{JKR}}^{\text{unloading}})=\delta_{\text{JKR}}^{\text{loading}} < 0$, $\delta_1 = \delta_2$, $a_{\text{PC}} = a_{\text{JKR}}$, and $P_{\text{PC}} = P_{\text{JKR}}$. Thus, the sliding portions (second integrals) in Eqs. \eqref{E:NewFormulation_Fy} and \eqref{E:NewFormulation_Ar} vanish, and Eqs. \eqref{E:NewFormulation_Fy} and \eqref{E:NewFormulation_Ar} reduce to Eqs. \eqref{E:Fy_Greenwood} and \eqref{E:Ar_Greenwood}, respectively.

\section{General behaviour of the model}
\begin{figure}[h!]
  \centering
  \includegraphics[width = \textwidth]{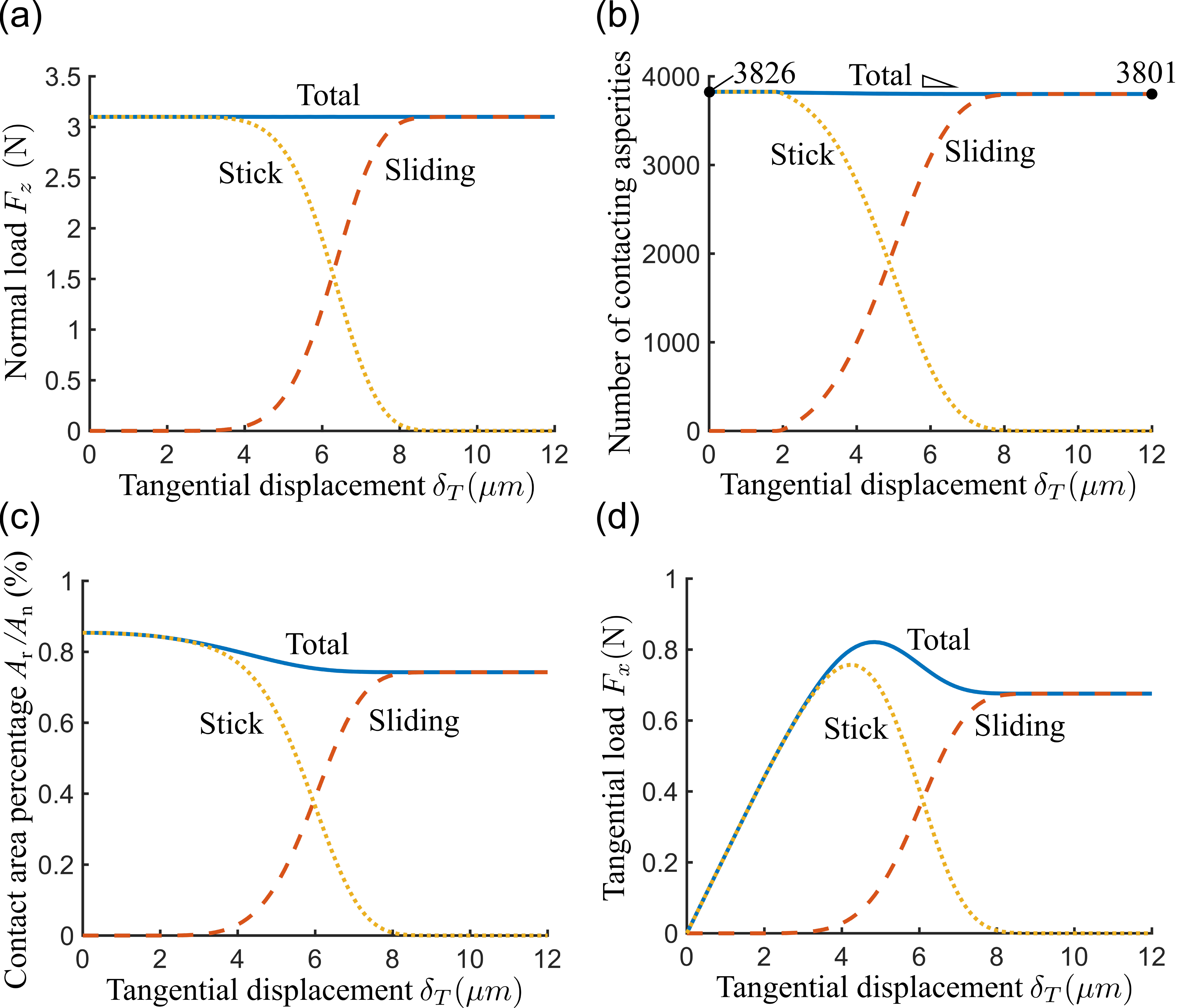}
  \caption{Typical behaviour of adhesive frictional contact between a rough elastomer and a rigid glass plate from the model. Macroscopic evolutions (solid line), as well as their stick (dotted line) and sliding (dashed line) contributions, against the tangential displacement $\delta_T$, of (a) normal load, (b) number of contacting asperities, (c) real contact area and (d) tangential load. The parameters (except for $\lambda$ and $\epsilon$) are of the same order of magnitude as those of the experiments from \cite{Sahli18}: $w = 40$ mJ/m$^2$, $E^* = 2.133$ MPa, $\lambda = 0.2$, $\tau_0 = 0.13$ MPa, $F_z = 3.1$ N, $A_n = 700$  mm$^2$, and $\epsilon = 0.2$ nm. The roughness-related inputs are $\eta = 652.1$ mm$^{-2}$, $\sigma_s = 0.032$ mm, and $R = 0.041$ mm. Tabor's parameter $\mu \approx 1217$.\label{fig:Fig_4}}
\end{figure}
The behaviour of the statistical model developed in Section \ref{sec:model} is illustrated using a reference set of parameters inspired by the experiments of \cite{Sahli18}, where a rough elastomer block in frictional contact with a glass plate is loaded tangentially until full sliding. The way the roughness parameters have been estimated is discussed in Appendix A. The elastic block is initially subjected to a purely normal load of $F_z = 3.1$ N, and is held constant for the rest of the test.

Even though the contributions to the normal load from the sticking and
sliding junctions monotonically decreases and increases, respectively, as the
tangential displacement, $\delta_T$, increases (see dotted and dashed lines
in Fig. \ref{fig:Fig_4}(a), which correspond to the contributions of the first and second integrals of Eq. \eqref{E:NewFormulation_Fy}, respectively), the prediction of normal load, $F_z$, by the
statistical model remains constant as imposed as a boundary condition. As implied by the status map in Fig. \ref{fig:Fig_2}(a), all contacting asperities being in the stick state gradually transit to the sliding state as the tangential displacement increases. This is confirmed by the dashed and dotted line in Fig. \ref{fig:Fig_4}(b), where the number of contacting asperities monotonically drops and increases within the sticking and sliding populations, respectively. Note that the variation of the total number of contacting asperities against the tangential displacement (see solid line in Fig. \ref{fig:Fig_4}(b)) is negligible, even though it decreases from $3826$ to $3801$. Such a loss of contacting asperities is caused by the jump-off-contact of asperities with negative indentation, see for example blue lines in Fig. \ref{fig:Fig_2}(b-d).

The real contact area, $A_r$, predicted by the statistical model shows a smooth drop until it converges to a constant when nearly all contacting asperities are in the sliding state, see solid line in Fig. \ref{fig:Fig_4}(c). This is consistent with the shear-induced area reduction phenomenon observed by Sahli \emph{et al}. \cite{Sahli18}, and it is due to the contact area shrinkage at single asperity level during the stick state. Clearly, as the contribution of the stick populations is nearly vanishing, the real contact area converges to its final value.

Figure \ref{fig:Fig_4}(d) presents the evolution of total tangential load, $F_x$, (solid line), which is also the sum of the contributions from the sticking and sliding populations. Initially, the tangential load almost linearly increases with the tangential displacement. This is because (1) nearly all contacting asperities remain stuck and (2) the tangential load at single asperity level increases almost linearly with the tangential displacement (see red line in Fig. \ref{fig:Fig_2}(d) as an example) when $\delta_T$ is relatively small (less than $2$ $\mu$m in Fig. \ref{fig:Fig_4}(d)). The tangential load enters a nonlinear increase stage, and is followed by a decrease until the tangential load equals $\tau_0 A_r$ once all asperities are in their sliding state. Most interestingly, a static friction peak (higher than the sliding friction) is reproduced by the statistical model. This is a common phenomenon in friction tests, and is frequently incorporated empirically in phenomenological friction laws. Here, the peak arises because the asperity-level model already presents a force drop (see Fig. \ref{fig:Fig_2}(d)) originating both from PC’s model, which yields a peak force of its own, and from an abrupt drop down to the sliding force proportional to Hertz’s contact area, at the sliding transition. Note that the tangential displacement at sliding inception (about $8$ $\mu$m) is several orders of magnitude lower than its experimental counterpart (about $1$ mm in \cite{Sahli18}). This is because the experimental displacement incorporates not only the interfacial displacement, but also the deformation of the tangential loading system with finite stiffness \cite{Papangelo20}.

\begin{figure}[h!]
  \centering
  \includegraphics[width = \textwidth]{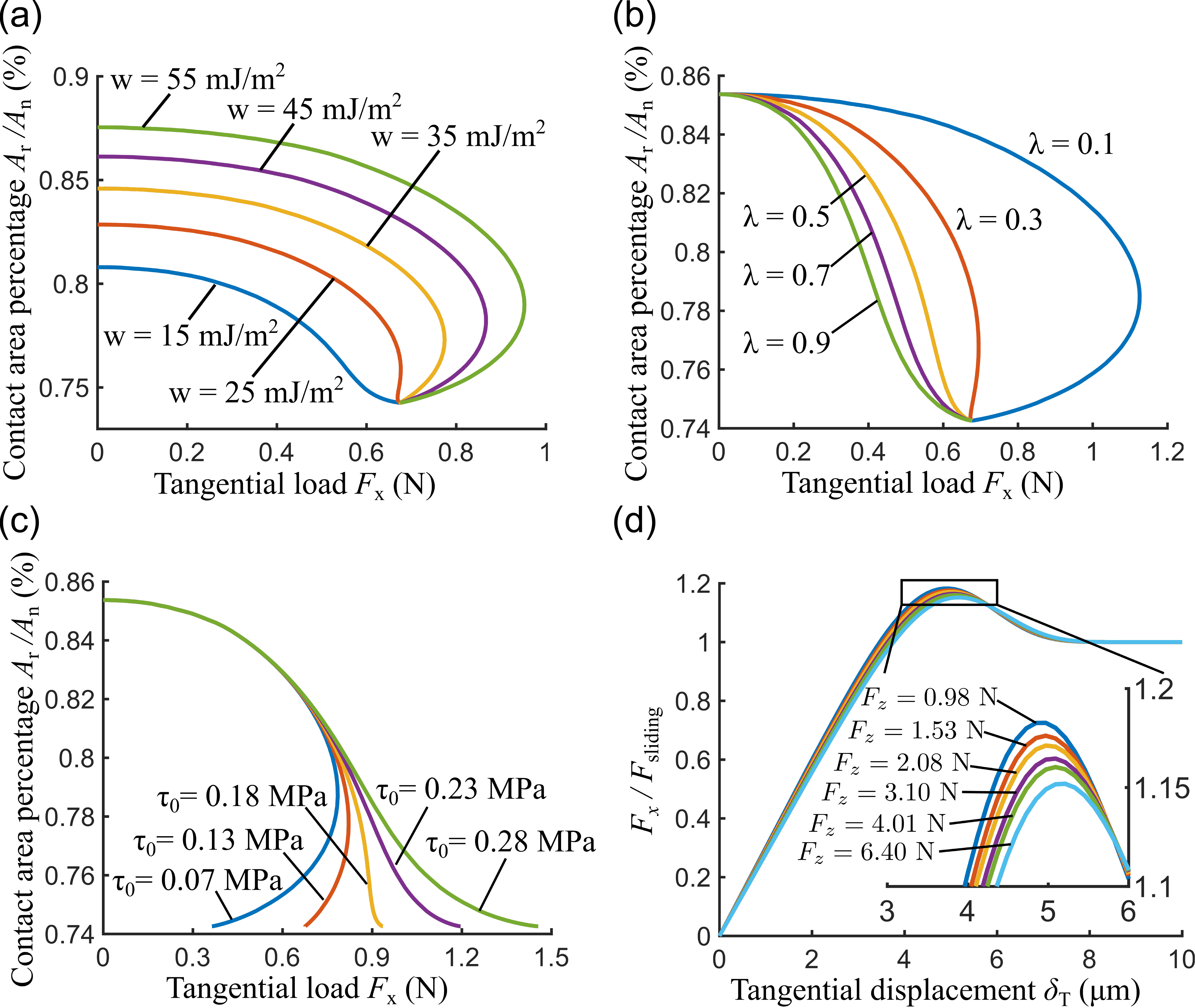}
  \caption{Effects of (a) work of adhesion $w$, (b) mode mixity $\lambda$, (c) frictional shear strength $\tau_0$ on the real contact area $A_r/A_n$ vs. tangential load $F_x$; (d) effect of normal load $F_z$ on the tangential load $F_x/F_{\text{sliding}}$ vs. tangential displacement $\delta_T$, as predicted by the statistical model. Each figure is achieved by varying one of the following base parameters: $w = 40$ mJ/m$^2$, $\lambda = 0.2$, $\tau_0 = 0.13$ MPa, $F_z = 3.1$ N. The other parameters are the same as that provided in the caption of Fig. \ref{fig:Fig_4}. The minimum Tabor's parameter $\min (\mu) \approx 633$.\label{fig:Fig_5}} 
\end{figure}

Figure \ref{fig:Fig_5} illustrates the effects of the work of adhesion, $w$, mode mixity parameter, $\lambda$, frictional shear strength, $\tau_0$, and normal load, $F_z$, on the real contact area and tangential load, especially the static friction peak. A reduction of the real area against the tangential displacement is observed for all studied cases. The work of adhesion and mode mixity parameter greatly affect the evolution of the tangential load. A larger work of adhesion means a tougher adhesive bond, and a lower mode mixity $\lambda$ implies a higher energy dissipation along the interface, which will also result in a tougher adhesive bond. Therefore, an obvious static friction peak can be observed in Fig. \ref{fig:Fig_5}(a) for $w \geq 25$ mJ/m$^2$ and in Fig. \ref{fig:Fig_5}(b) for $\lambda \leq 0.3$ (see the hook-like portion on the right-hand side of the curves). A relatively low work of adhesion and high mode mixity parameter result in an absence of static friction peak from the friction curves, e.g., $w = 15$ mJ/m$^2$ curve in Fig. \ref{fig:Fig_5}(a) and $\lambda = 0.9$ curve in Fig. \ref{fig:Fig_5}(b). The frictional shear strength $\tau_0$ within the sliding zone can also play an important role in determining the static friction peak. A relatively small $\tau_0$ can yield an obvious static friction peak, which may instead be buried in the tangential force curves as $\tau_0$ increases to a larger value, see Fig. \ref{fig:Fig_5}(c). The absence of a static friction peak was indeed observed in some friction tests (e.g., \cite{Weber19, Mergel19}), and it may be attributed to the combination of low surface energy, high mode mixity, and/or high frictional shear strength. Let $F_{\text{sliding}}$ be the tangential load acting on the rough surface during the sliding stage. The dimensionless tangential load $F_x/F_{\text{sliding}}$ vs. tangential displacement is given in Fig. \ref{fig:Fig_5}(d). As the inset shows, the relative difference between the static friction (the maximum of $F_x$) and the corresponding sliding friction reduces as the normal load increases. Some consistent empirical data may be found in \cite{Saad21} for silicon-silicon contact.

\section{Quantitative comparison with experiments}
\begin{figure}[h!]
  \centering
  \includegraphics[width = \textwidth]{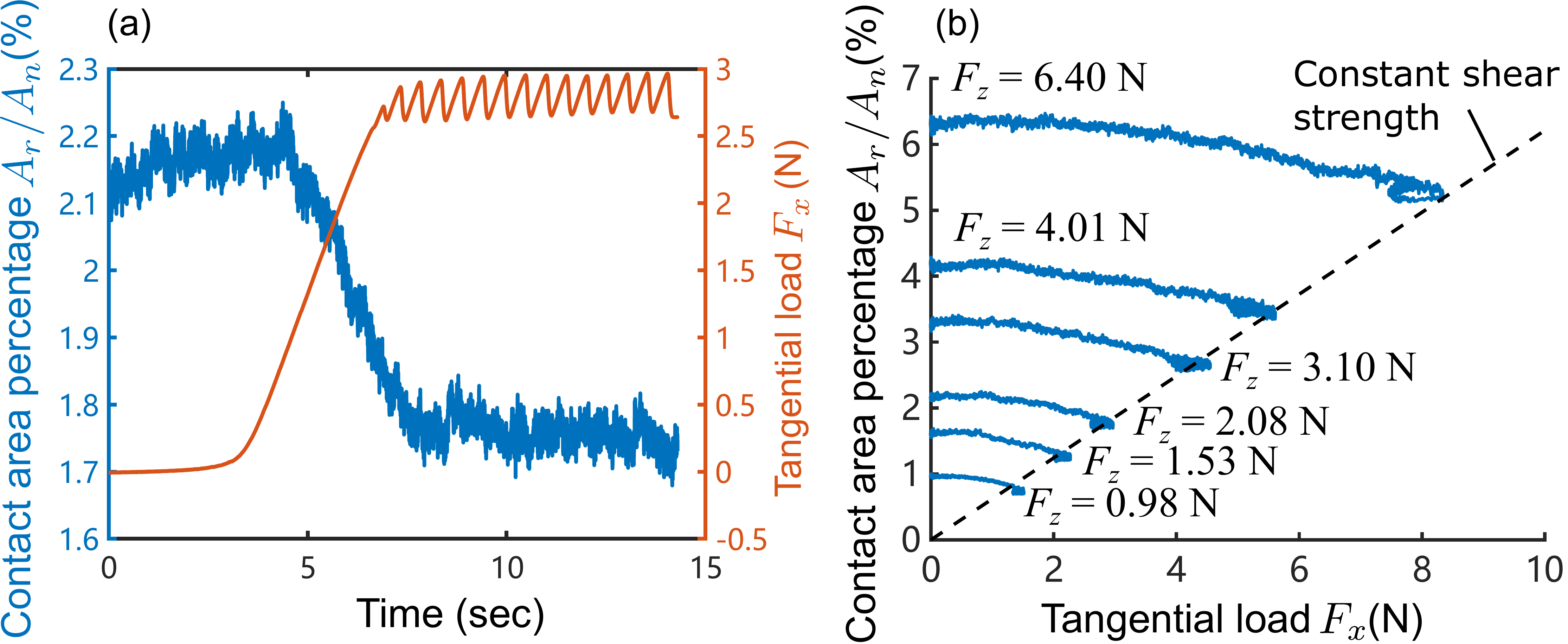}
  \caption{Selected adhesive friction test results measured by Sahli \emph{et al}., \cite{Sahli18}: (a) Evolution of real contact area and tangential load when $F_z = 2.08$ N; (b) Quadratic reduction of the fractional real contact area $A_r/A_n$ against the tangential load for all six normal loads. The tangential displacement is applied at a rate of $0.1$ mm/sec.\label{fig:Fig_6}}  
\end{figure}
Friction tests of a rough planar elastomer in contact with a smooth glass plate were conducted by Sahli \emph{et al}. \cite{Sahli18}, and some typical measurements are shown in Fig. \ref{fig:Fig_6}. In the experiments, the rough slider is driven tangentially through a thin steel wire with a stiffness of $9200$ N/m. At the beginning of the friction tests, the wire is fully unloaded, so that the motor needs to travel some distance before the wire actually exerts a force on the slider. Therefore, the first 3 seconds in Fig. \ref{fig:Fig_6}(a) are necessary for the wire to start applying any force on the slider, and the real contact area and the tangential load are nearly constant before that point. The real contact area is measured based on reflection images of the rough contact interface, acquired using a monochrome CCD camera. The pixels inside the real contact regions appear darker than those within out-of-contact regions. An automatically determined threshold value of grey level is used to segment the digital image into real contact area and out-of-contact area. Details about the image acquisition and segmentation algorithm can be found in the Supporting Information of Sahli \emph{et al}. \cite{Sahli18}. Periodic stick-slip cycles are found after the sliding inception, yielding a large number of data points accumulated at the tails of the friction curves $A_r(F_x)$ in Fig. \ref{fig:Fig_6}(b). This accumulation of quasi-cyclic data somewhat hides, especially at lower normal loads, the existence of an underlying hook-like shape of the curves like that predicted in the model (see Fig. \ref{fig:Fig_5}(a-c)), due to the existence of the static friction peak in the tangential force.

All the necessary input parameters of the statistical model are directly available from \cite{Sahli18} except for (1) the roughness-related parameters $R$, $\eta$ and $\sigma_s$, and (2) the mode mixity parameter $\lambda$. The roughness-related parameters may be estimated based on the topography data given in Appendix A. A relevant spatial resolution needs to be first identified so that the interaction and/or coalescence of  neighbouring micro-junctions can be accounted for in a effective way. As shown by Papangelo and Ciavarella \cite{Papangelo19a}, the $\lambda$ value can only be estimated by fitting the analytical solutions to the experimental results under shear. Even for smooth PDMS spheres prepared in the same group (with the same radius, material and tribometer), the fitted $\lambda$ values may differ by up to one order of magnitude \cite{Papangelo19a}(experimental data from Sahli \emph{et al}. \cite{Sahli18}: $\lambda = [5 - 6] \times 10^{-3}$; experimental data from Mergel \emph{et al}. \cite{Mergel19}: $\lambda = [7 \times 10^{-4} - 2 \times 10^{-3}$]). As a matter of fact, a two-step fitting procedure is adopted to determine the relevant spatial resolution and $\lambda$. In the first step, we consider the purely normal contact case in which the tangential displacement is zero. Since the results of purely normal contact are independent of $\lambda$, we focus on finding the spatial resolution such that the corresponding real contact area reaches a good agreement with the test results. In the second step, an optimal value of $\lambda$ is obtained so that the statistical model can recreate quantitatively the shear-induced reduction of the real contact area observed experimentally.

\begin{figure}[h!]
  \centering
  \includegraphics[width=\textwidth]{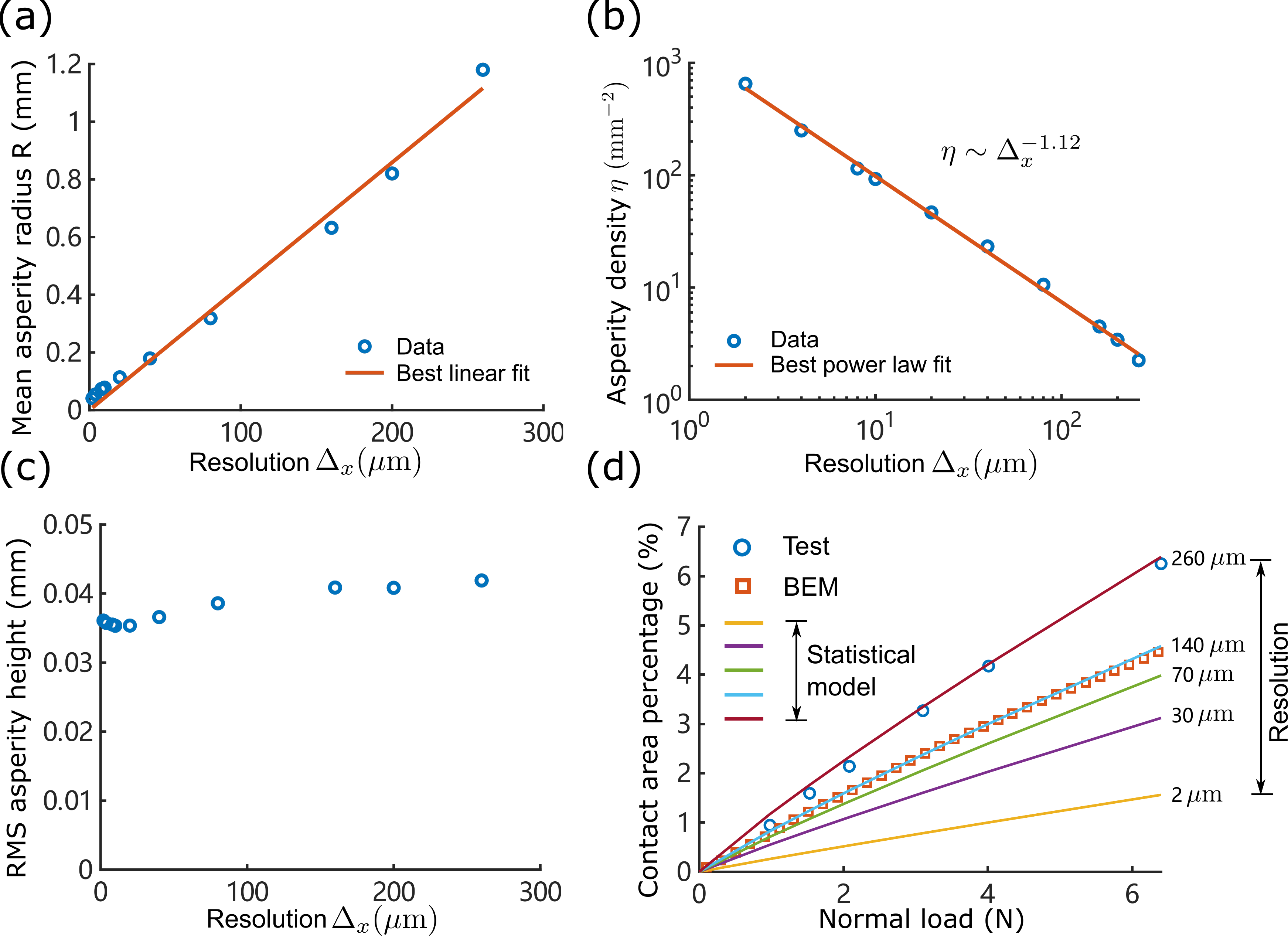}
  \caption{Evolution of (a) the mean asperity radius $R$, (b) the asperity density $\eta$, and (c) the root mean square (RMS) asperity height with the resolution; (d) comparison of the real contact area predicted by BEM and by the statistical model with the test data, under purely normal loading (i.e. with zero tangential displacement).\label{fig:Fig_12}}
\end{figure}

The topography data is measured on the rough mold used to create the rough elastomer surface, over a sampling area of $10 \times 10$ mm$^2$, which is one seventh of the entire nominal contact area between the glass plate and the elastomer slider. The raw topography data has a constant spatial resolution of $2$ $\mu$m. To explore how the roughness statistics vary with the spatial resolution, the measured topography data is re-sampled with various ratios to create multiple realizations at coarser scales, by considering successively one out of two, three, four, etc sampling points in $x$ and $y$ directions (yielding a spatial resolution of $\Delta_x = \Delta_y$ of $4$, $6$, $8$ $\mu$m, etc in both directions). The values of $R$, $\eta$ and $\sigma_s$ associated with each realization can then be evaluated following the approach given in Appendix A. As the resolution increases from $2$ $\mu$m to $260$ $\mu$m, the mean asperity radius $R$ increases linearly, see Fig. \ref{fig:Fig_12}(a); the asperity density $\eta$ drops as a power law of exponent $-1.12$, see Fig. \ref{fig:Fig_12}(b); the rms asperity height remains nearly constant, see Fig. \ref{fig:Fig_12}(c).

The statistical model, once solved using the roughness parameters extracted from the raw topography data (with a resolution of $2$ $\mu$m) significantly underestimates the real contact area, see yellow solid line in Fig. \ref{fig:Fig_12}(d). This is typical of statistical models based on independent asperities, like ours, because they neglect interaction and coalescence of neighboring micro-junctions \cite{Violano19a, Violano19b}. To test this explanation, a boundary element method (BEM) \cite{Pohrt14} is applied to predict the real contact area under purely normal loading (i.e. tangential displacement is zero). More details about this numerical model are given in Appendix B. The BEM model uses the full measured topography data as the input, and solves the linear elastic contact problem, including neighboring asperity interaction and micro-junction coalescence. The real contact area predicted by the BEM model (square) is indeed much closer to the experimental results (circle), compared with the statistical model (solid line, $2$ $\mu$m). The difference between the BEM calculations and experimental tests may be attributed to the fact that rough surface data used in the BEM is only one seventh of the entire nominal contact area, so that, the measured fraction may not be representative of the entire rough interface in the test. Indeed, M\"{u}ser \emph{et al.} have shown that the difference between the BEM and the experimental results can be negligible (see Fig. 10 in \cite{Muser17}), if the topography data of the entire contact interface is used as input for the BEM. They also found that statistical models of independent asperities significantly underestimate $A_r$, see Fig. 10 in \cite{Muser17}. This conclusion is consistent with Fig. \ref{fig:Fig_12}(d). Also note that, in future studies, a more accurate asperity-based rough surface contact model may be developed in the framework of ICHA [9, 49-51] which can explicitly consider neighbouring asperity interaction and micro-junction coalescence\footnote{Note that a way to implement neighboring asperity interactions in the Greenwood-Williamson framework would be to follow the works of Zhao and Chang \cite{Zhao2001} and Ciavarella \emph{et al}. \cite{Ciavarella08}. However, in those works, the neighboring asperity interaction is only achieved in an average sense, since an accurate account for the neighboring asperity interaction would need a detailed rough surface data, rather than a formless rough surface whose asperity heights follow the Gaussian distribution.}.

The competition between the increasing $R$ and decreasing $\eta$ with the spatial resolution results in an increasing real contact area $A_r$ predicted by the present statistical model, see Fig. \ref{fig:Fig_12}(d). The best agreement between the model prediction and the test results can be found when the resolution is approximately $260$ $\mu$m, see Fig. \ref{fig:Fig_12}(d). The corresponding roughness parameters are $R = 1.180$ mm, $\eta = 2.25$ mm$^{-2}$ and $\sigma_s = 0.0419$ mm.  Let us now discuss the physical relevance of such a coarse optimal resolution, based on the Power Spectrum Density (PSD) of the same rough topography given in Fig. S1 of \cite{Sahli19}. The PSD has a plateau at large scales followed by a power law at smaller scales. The transition point is associated with the roll-off wavelength of about $600$ $\mu$m. It is thus expected that the most visible surface asperities are the ones formed by the wavelengths larger than the roll-off wavelength (those present in the plateau of the PSD). Note that those wavelengths, according to Shannon’s theorem, can be accurately represented with our optimized spatial resolution of $260$ $\mu$m, because it is smaller than half the wavelength. The typical amplitude associated with those wavelengths is expected to be of the order of the rms roughness of the topography divided by $\sqrt{2}$, i.e. about $36$ $\mu$m. Thus, radii of curvature (estimated as that at the summits of a sine wave with those wavelength and amplitude) larger than $250$ $\mu$m are expected, suggesting that a mean asperity radius of $1.180$ mm may very well be representative of the most relevant summits. Note, that the pixel size in the contact area images ($25$ $\mu$m) is sufficiently smaller than the optimized resolution ($260$ $\mu$m), so the contact images are expected to be able to resolve the micro-junctions formed by those larger asperities. The spectral contents with the wavelengths lower than the roll-off wavelength may only create a ``protuberance on protuberance" structure, presumably hardly detectable by the optical method of Sahli \emph{et al}. \cite{Sahli18}.

Choosing $260$ $\mu$m as the relevant effective resolution, we are now ready to find the optimized $\lambda$ so that the $A_r(F_x)$ curves predicted by the present model can have good agreement with the test results given in Fig. \ref{fig:Fig_6}(a). $\lambda = 0.0023$ best fitted from the single asperity contact test \cite{Papangelo19a} may be used as an initial guess. However, such a small $\lambda$ value causes a very small area reduction rate, in particular a very flat initial part of the $A_r(F_x)$ curve. According to Fig. \ref{fig:Fig_5}(b), the reduction rate can be increased by larger $\lambda$. A higher value of $\lambda$ in the rough surface scenario may be explained as follows: in the above-mentioned asperity contact test, the spherical asperity and glass plate are optically smooth, while the rough surface on the elastomer slider has a topography spanning a wide range of wavelength, including wavelengths (a few $\mu$m) much smaller than the micro-junctions that our model is focusing on (typically hundreds of $\mu$m diameter). Thus, there exists a small roughness on the main asperities forming those micro-junctions. Although the physical ingredients affecting the value of $\lambda$ are far from being understood, we can speculate that such a small scale roughness reduces the interfacial dissipation under shear, compared to a perfectly smooth asperity. Such a reduction of the dissipation would certainly increase the value of $\lambda$, by an extent that is not predictable in the current state of knowledge in the field.

\begin{figure}[h!]
  \centering
  \includegraphics[width=0.7\textwidth]{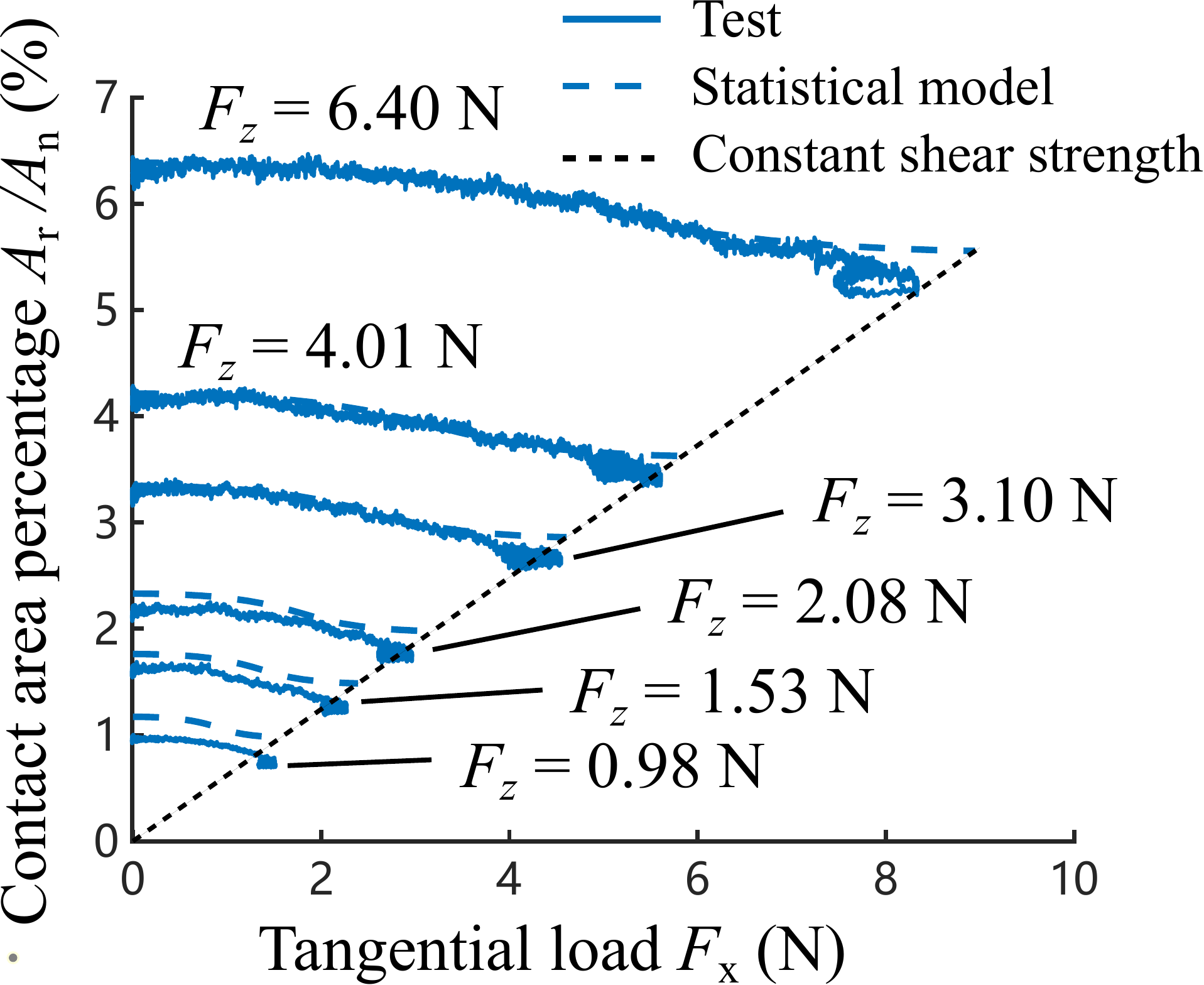}
  \caption{Comparison of the fractional real contact area $A_r/A_n$ vs. tangential load $F_x$ under six different normal loads measured by Sahli \emph{et al}. \cite{Sahli18} (solid curves) and predicted by the statistical model (coarse dashed curves) and a constant shear strength model (fine dashed line). Choosing the optimized resolution of $260$ $\mu$m, the corresponding roughness parameters are: $R = 1.180$ mm, $\eta = 2.25$ mm$^{-2}$ and $\sigma_s = 0.0419$ mm. $\lambda = 0.03$. The other parameters are the same as that used in Fig. \ref{fig:Fig_5}. Tabor's parameter $\mu \approx 1939$.\label{fig:Fig_13}}
\end{figure}

The best fit of $A_r(F_x)$ predicted by the statistical model to the test results can be found with $\lambda = 0.03$. The corresponding $A_r(F_x)$ curves for all six normal load cases along with the test results are shown in Fig. \ref{fig:Fig_13}. The statistical model reproduces a convex-shaped reduction of the real contact area against the tangential load, very similar to the quadratic-like reduction observed experimentally by Sahli \emph{et al}. \cite{Sahli18}, a feature that the previous model of \cite{Scheibert20} could not capture. A better agreement is found at higher normal loads, due to a better accuracy of the predicted real contact area when tangential load is zero, see Fig. \ref{fig:Fig_12}(d). The hook-like portions at the right end of $A_r(F_x)$ curves can hardly be reproduced at the same time because, in the experiments, it is presumably an unstable dynamic feature caused by stick-slip. A better solution for fixing this mismatch might be to use the dynamic formulation proposed by Scheibert \emph{et al.} \cite{Scheibert20}.

We are now in a position to comment on the previous model by Scheibert \emph{et al}. \cite{Scheibert20}, already mentioned in the Introduction. They developed an asperity-based dynamic contact model in which the real contact area is composed of multiple elliptic contact spots. The initial areas of contact of individual micro-junctions at zero tangential load are directly taken from the optical measurement of \cite{Sahli18}. The evolution of each contact spot under shear is governed by a quadratic phenomenological law:
\begin{equation}
A_i = A_{0i} - \alpha_b \frac{1}{A_{0i}^p} f_i^2,
\end{equation}
where $A_{0i}$ and $A_i$ are the initial and instantaneous areas of the $i^{\text{th}}$ elliptical contact spot, respectively. The corresponding tangential force at each contact spot, $f_i$, is determined based on the tangential stiffness of an elliptical contact in the full stick condition, until it converges to the sliding friction $f_i = A_{i} \tau_0$. The predicted dynamic friction has excellent agreement with the test results, not only before the sliding inception, but also during the stick-slip stage. In contrast, instead of a quadratic shape, the predicted reduction of real contact area shows a nearly linear trend against the tangential force. According to the discussion given by Scheibert \emph{et al}. \cite{Scheibert20}, their disagreement may be attributed to the following two reasons:
\begin{enumerate}
  \item Constant fitting parameters, $p$ and $\alpha_b$, in the phenomenological law;
  \item Lack of elastic coupling between neighboring micro-junctions.
\end{enumerate}
The first reason actually suggests that the empirical asperity scale law they have used is not fully satisfactory. Indeed, a detailed expression of $\alpha_b/A_{0i}^p$ was derived by Papangelo and Ciavarella, see Eqs. (27--29) in \cite{Papangelo19a}, based on the PC model, which appears to be an alternative, more physically-based contact mechanics law. As a matter of fact, according to the results of our statistical model, the quadratic shape of contact area reduction is recovered if the PC model is used to describe the shear-induced area reduction, see Fig. \ref{fig:Fig_13}. This is a strong signal that, if a more sophisticated form of $p$ and $\alpha_b$ were introduced, the model developed by Scheibert \emph{et al}. might better predict the shape of the area reduction. Additionally, the second invoked reason may not be as paramount as the first one since the present statistical model also does not consider the coupling between neighboring asperities and still behaves better.

\section{Conclusions}
The adhesive friction behaviour of a linear elastic rough contact interface is modeled using a statistical model of independent spherical asperities. The coupling between friction and adhesion at the asperity level is described using Papangelo and Ciavarella’s model, in which the periphery of each asperity is considered as an interfacial crack under a mixed mode loading. As tangential displacement increases, more and more micro-junctions switch from a sticking to a slipping state, until global sliding occurs. The statistical model satisfactorily reproduces the phenomenon of shear-induced reduction of the real contact area observed in recent friction tests, which is attributed to junction shrinkage at the asperity level. Remarkably, our model also reproduces a static friction peak at the macroscale without explicitly introducing two different friction levels (static and dynamic/sliding). It also demonstrates how the static friction peak and contact area evolution depend on the normal load and the interfacial properties such as surface energy, mode mixity and frictional shear strength. Generally, a ``tougher" interface (larger surface energy and smaller mode mixity parameter) results in more contact area and a more pronounced static friction peak, all the more so as the normal load is reduced. The static friction peak can be made to ``disappear" due to a relatively higher frictional shear strength and a ``weaker" interface - again a behaviour which has been observed in experiments. The prediction of the friction force and real contact area quantitatively matches the experimental measurements, if a specific evaluation procedure is followed to estimate the relevant roughness parameters and the mode-mixity parameter. Overall, this work provides important insights about how key microscale properties (e.g. surface, mode mixity etc.) operating at the asperity level can combine with the surface statistics to simulate some of the most relevant macroscopic responses observed in experiments on the friction of rough elastic surfaces - most notably, the evolution of friction and real contact area during the transition from static to sliding friction. The paper, therefore, offers important fundamental insights into the mechanism of friction itself. In the future, those insights might be a useful aid in designing engineered friction systems (i.e. functionalized textured interfaces etc). The work also paves the way for the development of more comprehensive asperity-based and full continuum models for the adhesive friction of rough elastic contacts.

\section*{Acknowledgements}
This work is supported by the Leverhulme Trust through Project Grant ``Fundamental mechanical behavior of nano and micro structured interfaces" (RPG-2017-353), the National Natural Science Foundation of China (Grant No. 52105179), and LABEX MANUTECH-SISE (ANR-10-LABX-0075) of Universit\'{e} de Lyon, within the program Investissements d’Avenir (ANR-11-IDEX-0007) operated by the French National Research Agency (ANR). JS is indebted to Institut Carnot Ing\'{e}nierie@Lyon for support and funding.

\appendix

\setcounter{figure}{0}
\setcounter{equation}{0}
\renewcommand{\thefigure}{A.\arabic{figure}}
\renewcommand{\theequation}{A.\arabic{equation}}
\renewcommand{\thetable}{A.\arabic{table}}

\section*{Appendix A. Statistics of measured rough surface topography}
\begin{figure}[h!]
  \centering
  \includegraphics[width=\textwidth]{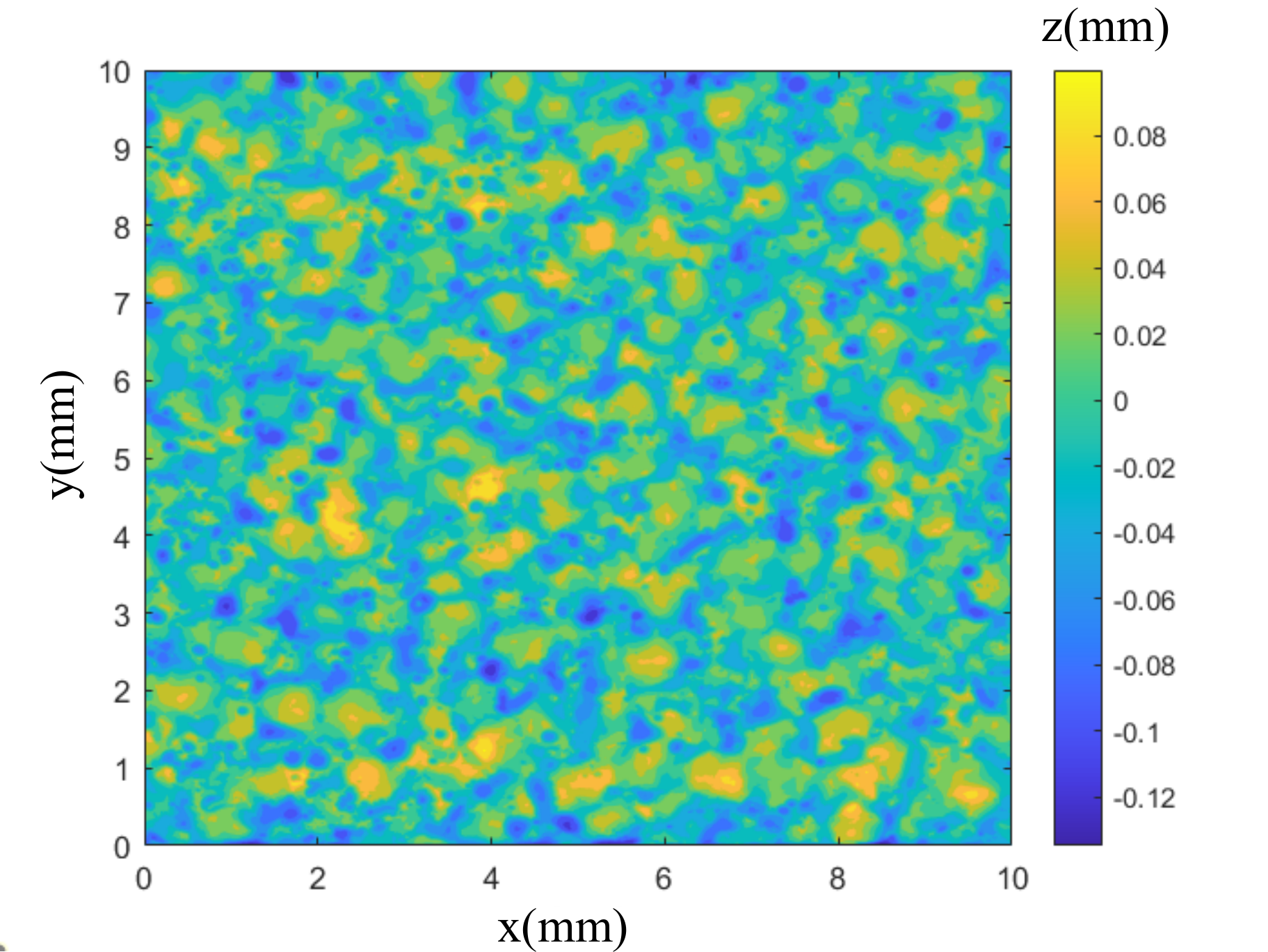}
  \caption{Map of the measured rough surface topography \cite{Sahli18, Sahli19}. The rough surface consists of $5001 \times 5001$ sampling points over a sampling area of $10 \times 10~\text{mm}^2$. The rough surface is leveled and the mean surface height is zero.\label{fig:Fig_A1}}
\end{figure}

The rough surface used in this study was measured by Sahli \emph{et al}. \cite{Sahli18, Sahli19} which is one seventh of the nominal contact area of the elastomer block ($35 \times 20~\text{mm}^2$) used in the friction test, see the rough surface height map in Fig. \ref{fig:Fig_A1}. The rough surface is sampled with a resolution of 2 $\mu$m.

Three roughness-related parameters, namely, asperity density, $\eta$, mean asperity radius, $R$, and standard deviation of the asperity height ,$\sigma_s$, are needed for the statistical model developed in Section \ref{sec:model}. The asperities are identified using the method of $4$-nearest-neighbors, i.e., a sampling point is identified as the summit of an asperity if it is higher than the nearest four sampling points. The asperity density over the sampling area $A_n = 10 \times 10~\text{mm}^2$ is $\eta = N/A_n = 652.1$ mm$^{-2}$ where $N$ is the number of asperities. The standard deviation of asperity height is $\sigma_s = \sqrt{\langle z^2 \rangle} = 0.032$ mm where $z$ is the summit height of all asperities measured about the mean summit height level, i.e., $\langle z \rangle = 0$. The mean asperity radius $R = 0.041$ mm is the average of the summit radius of the asperities $R = \frac{1}{2} \langle |\partial z^2/\partial^2 x|^{-1} + |\partial z^2/\partial^2 y|^{-1} \rangle$ which can be approximated using central differentiation. The above calculations of $R$, $\eta$ and $\sigma_s$ are associated with the resolution of 2 $\mu$m. Analogous measurements based on re-sampled data yielding coarser spatial resolutions are given in Fig. \ref{fig:Fig_12}.

\setcounter{figure}{0}
\setcounter{equation}{0}
\renewcommand{\thefigure}{B.\arabic{figure}}
\renewcommand{\theequation}{B.\arabic{equation}}
\section*{Appendix B. Boundary element method for adhesive contact}
The Pohrt \& Popov's model \cite{Pohrt15} is an adhesive contact model applied to the pre-loading stage of the adhesive friction where $F_x = 0$ N and adhesion is included. Consider a rough elastic half-space in purely normal non-adhesive contact with a rigid flat, see Fig. \ref{fig:Fig_3} where $F_x = 0$ N. The geometrical relation at the contact interface has the following form
\begin{equation}\label{E:BEM_eq1}
  g(x, y) = u_z(x, y) - h(x, y) + d ~~~ (x, y) \in \Omega,
\end{equation}
where $g$ is the interfacial gap; $\Omega$ is the computational domain; $u_z$ is the normal surface deflection which can be determined by the convolution between the Boussinesq solution and contact pressure $p$ \cite{Johnson87}; $d$ is a load-dependent distance. The solutions, $p(x, y)$ and $g(x, y)$, must satisfy the following boundary conditions
\begin{align}
&g(x, y) = 0 ~~~~~~~~~~~~~~~~~~~~~~ \text{where} ~~~ (x, y) \in \Omega_c, \label{E:BEM_eq2}\\
&g(x, y) > 0, ~~~ p(x, y) = 0 ~~~~ \text{where} ~~~ (x, y) \in \Omega_{nc}, \label{E:BEM_eq3}
\end{align}
where $\Omega_c$ and $\Omega_{nc}$ are the contact and out-of-contact regions, respectively. Finally, the load equilibrium in the normal direction must be maintained
\begin{equation}
F_z = \iint_{\Omega_c} p(x, y) dx dy \label{E:BEM_eq4}.
\end{equation}
Since the measured rough topography of a finite size $10 \times 10$ mm$^2$ is only one seventh of the entire nominal contact area $35 \times 20$ mm$^2$ used in the friction tests, the computational domain $\Omega$ is assumed to be periodic. The load-dependent $d$ reduces to the average interfacial gap. The entire measured rough topography shown in Fig. \ref{fig:Fig_A1} is used in the BEM model, where the size of each surface element is $2 \times 2$ $\mu$m.  Fast Fourier Transform (FFT) is adopted to accelerate the calculation of the normal deflection $w$ \cite{Polonsky99, Xu17}. For a non-adhesive contact, the solutions of nonlinear equations, Eqs. (\ref{E:BEM_eq1}--\ref{E:BEM_eq4}), are iteratively solved using the conjugate gradient (CG) method proposed by Polonsky and Keer \cite{Polonsky99, Xu17}.

For a non-adhesive contact, the BEM model can be divided into two parts, a normal contact BEM solver and a mesh-dependent delamination criterion. The former part solves Eqs. (\ref{E:BEM_eq1}--\ref{E:BEM_eq4}) iteratively with a given real contact area $\Omega_c$ using the conjugate gradient (CG) method \cite{Polonsky99, Pohrt14}. In the latter part, the contacting interface is considered as an interface crack. The surface elements are delaminated where the normal tractions $p(x, y)$ violate the Griffith criterion, which states that the crack propagates when the strain energy release rate (G) is larger than $w = 27$ mJ/m$^2$. Pohrt and Popov \cite{Pohrt15} developed an analytical solution of $G$ using the Boussinesq solutions, and their derivation results in a mesh-dependent form
\cite{Pohrt15}:
\begin{equation}\label{eq:PohrtPopov_G}
G = \frac{p^2 \chi}{\Delta_x \Delta_y E^*},
\end{equation}
where
\[
\chi = \frac{1}{3 \pi}\left( \Delta_x^3 + \Delta_y^3 - \Delta_x^2 \bar{\Delta} - \Delta_y^2 \bar{\Delta} \right) +
       \frac{1}{2 \pi}\Delta_x \Delta_y\left[
       \Delta_x \log\left( \frac{\bar{\Delta} + \Delta_y}{\bar{\Delta} - \Delta_y} \right) +
       \Delta_y \log\left( \frac{\bar{\Delta} + \Delta_x}{\bar{\Delta} - \Delta_x} \right)
       \right].
\]
$\Delta_x = 2 ~\mu$m and $\Delta_y = 2 ~\mu$m are mesh sizes in the $x$ and $y$ directions, respectively; $\bar{\Delta} = \sqrt{\Delta_x^2 + \Delta_y^2}$. The analytical solution $G$ in Eq. \eqref{eq:PohrtPopov_G} quantifies the released strain energy due to the delamination of a rectangular element of unit area. Substituting Eq. \eqref{eq:PohrtPopov_G} into $G > w$, the energy-based delamination criterion can be rewritten as a normal traction-based one, i.e., $p < -\Sigma$ where
\begin{equation}\label{eq:max_sigma}
\Sigma = \sqrt{\frac{\Delta_x \Delta_y w E^*}{\chi}}.
\end{equation}

\end{document}